\documentclass[
amsmath,%
aps,%
final,%
notitlepage,%
oneside,%
twocolumn,%
nobibnotes,%
nofootinbib,
superscriptaddress,%
noshowpacs,%
]%
{revtex4}

  \usepackage[cp1251]{inputenc}
  \usepackage[dvips]{graphicx}

\newlength{\boxw}
\newlength{\boxh}
\newcommand{\rf}[1]{(\ref{#1})}

\newcommand{\bc}{\begin{center}}
\newcommand{\ec}{\end{center}}
\newcommand{\be}{\begin{eqnarray}}
\newcommand{\ee}{\end{eqnarray}}

\newcommand{\bfr}{\begin{flushright}}
\newcommand{\efr}{\end{flushright}}
\newcommand{\bfl}{\begin{flushleft}}
\newcommand{\efl}{\end{flushleft}}
\newcommand{\dsp}{\displaystyle}

\newcommand{\ra}{{\rho^{a}}}

\newcommand{\rft}{\rho_f}

\newcommand{\gaf}{g_{af}}

\newcommand{\Ha}{H_{a}}
\newcommand{\Hf}{H_{f}}

\newcommand{\Vaf}{V_{af}}

\newcommand{\Vafo}{\tilde{V}_{af}}

\newcommand{\Tr}{\mbox{Tr}}

\newcommand{\bk}{{\bf k}}
\newcommand{\bkl}{{\bf k}\lambda}

\newcommand{\wk}{{\omega_k}}

\renewcommand{\ap}{\hat{a}_{\bkl}^{+}}
\newcommand{\am}{\hat{a}_{\bkl}}
\newcommand{\kl}{\textbf{k}\lambda}
\newcommand{\Qind}{Q^{J_{j_1}M_{j_1} J_{j_2}M_{j_2}}_{J_d M_d \sigma \sigma^{'}}}
\newcommand{\Bind}{B^{J_{j_1}M_{j_1} J_{j_2}M_{j_2}}_{J_d M_d \sigma \sigma^{'}}}

\newcommand{\Spa}{^{a}}

\begin{document}

\title{Quantum interference effects in degenerate systems.
\\ Spontaneous and stimulated radiation.}

\author{A.~A. Panteleev, Vl.~K. Roerich}
\affiliation{FSUE УSRC RF TRINITIФ, Federal State unitary Enterprise УState Research Center of Russian
Federation Troitsk Institute for Innovation and Fusion ResearchФ, TRINITI, Troitsk, Moscow region, Russia
142190}\email[Email address: ]{vroerich@triniti.ru}

\begin{abstract}

We study the effect of quantum interference on the structure and properties of spontaneous and stimulated
transitions in a degenerate $V$-type three-level atom with an arbitrary total momentum of each state.
Explicit expressions for the factors in the terms of the relaxation operator and stimulated transition
operator with account of quantum interference effects are obtained. It has been demonstrated that the
condition for the dipole transition moments to be parallel is insufficient and for conventional atoms the
interference cross terms are zeros for both operators. The conditions when quantum interference influences
the properties of the relaxation operators have been analysed. For the stimulated transitions this
condition is shown to be anisotropy of the photon field interacting with the atoms. This case is minutely
studied for the $D$-line of alkali metals.
\end{abstract}

\maketitle

\section{ Introduction.}

Analysis of spontaneous decay of excited atomic states and spectral-angular, polarization, spatial, and
other characteristics of scattered radiation has long been the major focus of research in quantum optics.

Recent studies have substantially raised attention to the phenomenon of quantum interference occurring at
spontaneous emission from two or more closely separated neighbor states. Such a trend is explained by the
fact that these interference processes result in a range of various effects which are likely to have
important practical applications. Quantum interference at spontaneous emission is the interference
phenomenon of Fano type \cite{Fano} and is due to the fact that the states decay to the common continuum.
In \cite{Fano} it was demonstrated that in such a case the relaxation operator contains off-diagonal
terms. In \cite{Devdar} the off-diagonal terms in the relaxation operator were obtained in a similar form
while solving the problem of crossed quasi-steady states interaction with continuous spectrum. A similar
operator was derived in \cite{Dalid} to describe exchange interaction for single-nucleon quasi-steady
states.

The recent interest towards interference effects in spontaneous
emission arose in connection with the works of Harris and Imamoglu
\cite{5,6,7} where they demonstrated a possibility of  lasing
without inversion when two excited states interact with common
quantized radiation field. The subsequent changes in absorption
and amplification profiles were explained by the off-diagonal
terms in the relaxation operator. Lasing without inversion due to
quantum interference in a $V$-type three-level system was also
studied in \cite{8,9,10}. It must be noted that a significant
change in the profile of absorption and amplification is as well
possible for two-level systems in a strong field. In this case
there can be amplification of resonant photons near the
frequencies detuned from the central peak by the value of Rabi
frequency \cite{p11,p12,p17,p30,p31,p32}. This phenomenon was
experimentally observed in the radiofrequency range \cite{p40} and
in optical range as reported in \cite{p41}. It is also worth
noting that lasing without inversion was also discussed, for
instance, in \cite{Kochar,Kochar1,Kochar2} when treating the
three-level system of $\Lambda$-type.

In further studies of quantum interference a great number of other effects were predicted. The prime
phenomenon to mark out is the effect of spectral line narrowing. Zhou and Swain were the first to show
this effect while they were studying the spontaneous decay of a $V$-type three-level atom. This effect as
well as observation of so-called dark states were analysed in
\cite{11,quench-zhu,quench-scully,quench-scully1,quench-scully2,quench-scully3,quench-scully4,Roerich2000}.
In \cite{quench-scully1,1,2} they demonstrated the possibility for quenching of spontaneous emission. In
\cite{exper} this phenomenon was experimentally verified for molecular transitions in sodium dimer
systems. The initiation of quantum beats during spontaneous decay was studied in
\cite{qbeat-hegerfeldt,qbeat-hegerfeldt2,qbeat-hegerfeldt1,qbeat-patnaik,qbeat-patnaik1}. Quantum
interference was studied for the cases of both monochromatic and biharmonic external fields. The
conditions for coherent population trapping in a biharmonic laser field were considered in \cite{2,3,4}.
Besides, in \cite{phase-dep-martinez,phase-dep-knight,phase-dep-knight1,phase-dep-knight2} it was
demonstrated that for such systems the line shape is highly dependable on the phase difference of the
driving fields. In \cite{phase-dep-knight,phase-dep-knight1,phase-dep-knight2} they studied the conditions
for spontaneous intensity control by means of variation of laser field phase and phase switch-on time.

At least two correlated paths are required for quantum interference to occur. The two elementary systems
to provide these conditions are the nondegenerate three-level atoms of $V$ and $\Lambda$ types for which
all of the above mentioned effects are already true. By now this has mainly been the model used to study
quantum interference. However, some modifications have also been introduced, i.e., the system was studied
with some additional levels in various configurations, several driving fields, etc. Yet, it must be noted
that such a system is nothing but a model and does not reflect some properties of real physical systems
being of great importance for practical implementations and likely to change essentially the consideration
of the quantum interference itself. Some drawbacks of this model are the facts that it does not take any
account for possible level degeneration, neglects angular and polarization distribution of both driving
and spontaneous fields. In the papers we referred to above the quantum interference was conditioned by
nonorthogonal (parallel ideally) dipole transition vectors \cite{aniso-gsa}, \cite{quench-zhu},
\cite{quench-berman}, \cite{qbeat-patnaik}, \cite{qbeat-patnaik1}. However, as we show it in this paper
below this condition is insufficient if a rather full-scale model is considered.

The aim of this work is to study the effect of quantum
interference upon radiative relaxation for real atomic transitions
with account for level degeneracy in the magnetic moment
projection. In this work we fully focus on the properties of the
$V$-type model for both spontaneous and stimulated transitions
allowing for the hyperfine splitting. The $\Lambda$-type model can
be as well studied by means of the techniques described below.

This work is a sequel of our earlier series of papers \cite{Pant1997,Roer2000.2,Roer2000.3}, devoted to
studying the properties of spontaneous emission of an atom exposed to a strong monochromatic field of
arbitrary polarization composition. This paper is structured as follows. In the section \ref{bas} we state
the problem and derive the basic equations. Section \ref{Spon} is to study spontaneous processes. Section
\ref{Stim} is devoted to stimulated phenomena. We derive the operator to describe transitions stimulated
by a non-coherent field and consider thoroughly the cases of homogeneous and heterogeneous fields for the
$D$-line of alkaline metals. In the appendix we derive the relaxation operator subject to the hyperfine
level splitting.

\section{\label{bas} Basic equations.}
We consider a quantized electromagnetic field interacting with a
degenerate $V$-type three-level atom with $b\to d$, $c\to d$
(see~Fig.~\rf{fig:1}) allowed transitions having the total angular
momentum $J_j$, $j=b,c,d$. Using the dipole approximation, the
Hamiltonian (in units $1/\hbar$) of the system can be written as a
sum:
\begin{equation}
 \label{Gamilton}
  \hat{H}=\Ha+\sum_{\kl}\Hf+\sum_{\kl}\Vaf.
\end{equation}
Here, in (\ref{Gamilton}) $\Ha$ describes an unperturbed atomic
system:
\begin{equation}
 \label{GamH0}
  \Ha=\sum_{j=b,c}\sum_{M_j}\frac{\epsilon(J_j,M_j)}{\hbar}|j,M_j\rangle\langle M_j,j|,
\end{equation}
\begin{figure}
    \includegraphics[scale=0.48]{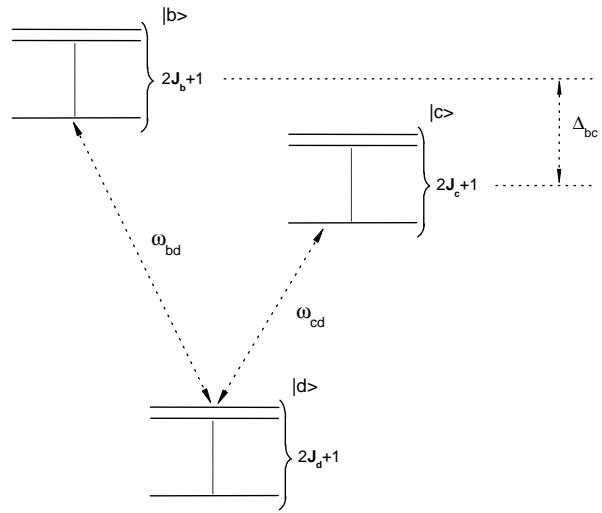}
   \caption{\small Degenerate three-level atom.}
  \label{fig:1}
  \end{figure}
where $\epsilon(J_j,M)$ is the energy of an atomic state with the total angular momentum $J_j$ and its
projection $M_j$ on the quantization axis (z), while $\langle M_j,j|$ is the corresponding vector for this
state. Since we further neglect splitting of the magnetic sublevels as well as other effects resulting
into magnetic moment projection dependence of the energy, we write:
\begin{equation}
 \label{GamH0end}
  \Ha=\sum_{j=b,c}\sum_{M_j}\omega_{jd}|j,M_j\rangle \langle M_j,j|,
\end{equation}
where $\omega_{jd}$ is the $j\to d$ transition frequency for any
pair of sublevels.

The second term in (\ref{Gamilton}) describes the Hamiltonian of
the quantized field:
\begin{equation}
 \label{GamHf}
  \hat{H_f}=\wk\ap\am,
\end{equation}
with $\wk$ being the photonic frequency with  wave vector $\bk$, $k=|\bk|$ and polarization
$\lambda=\pm1$, $\ap$ and $\am$ are the creation and annihilation operators for the  photons respectively.

The atom-field interaction is described by the expression:
\begin{equation}
 \label{GamVq}
 \begin{array}{c}
  \dsp{\Vaf=i\sum_{j=b,c}\sum_{m_j m_d}}
  \dsp{  g^{{\bf k}\lambda}_{jd}(M_j M_d)\hat{a}_{{\bf k}\lambda}|j M_j\rangle \langle M_d d| }+ H.c \\
 \end{array}
\end{equation}
where
 \be
   \label{const}
   \dsp{g^{{\bf k}\lambda}_{jd}(M_j M_d)=\sqrt{\frac{2\pi\wk}{\hbar
   W}}(\vec{\mu}_{jd}(M_j M_d),
  \epsilon(\bk,\lambda)) }
 \ee
is the coupling constant with $\vec{\mu}_{jd}(M_j M_d)$ being the
dipole transition operator. $\epsilon(\bk,\lambda)$ is the unitary
polarization vector, and $W$ denotes the quantization volume.
Using the Wigner-Eckart theorem \cite{Varshal} the coupling
constant \rf{const} can be written as
 \be
  \label{const2}
  g^{{\bf k}\lambda}_{jd}(M_j M_d)= \frac{C^{J_j M_j}_{ J_d M_d 1\sigma}}{\sqrt{2J_j+1}} ||\mu_{jd}||
   \sqrt{\frac{2\pi\wk}{\hbar W}}D_{\lambda\sigma}^1.
 \ee
where  $C^{J_j M_j}_{ J_d M_d 1\sigma}$ are the Glebsch-Gordan coefficients,
$D_{\lambda\sigma}^1=D_{\lambda \sigma}^1(\alpha=\phi,\beta=\theta,\gamma=0)$ is the Wigner function
\cite{Landau3}, with the angles $\phi,\theta$ determining the direction of the wave vector ${\bf k}$ of an
emitted photon to the quantization axis (z). It must be noted that due to the properties of the
Glebsch-Gordan coefficients $g^{{\bf k}\lambda}_{jd}(M_j M_d)$ is different from zero only if
$\sigma=M_j-M_d$, while $\sigma$ can be $0,\pm1$ because of the selection rules.

It is now advantageous for us to change to the interaction picture
where the Hamiltonian has the form:
 \be
 \label{Gamilton1}
  \begin{array}{c}
  \hat{H_i}=\Vafo, \;\;\;\;\;
  \dsp{\Vafo(t)=i\sum_{j=b,c}\sum_{m_j m_d}} \\
  \dsp{  g^{{\bf k}\lambda}_{jd}(M_j M_d)\am|j M_j\rangle \langle M_d d|e^{i(\omega_{jd}-\wk)t}+ H.c} \\
  \end{array}
 \ee

We further treat the system using the cluster expansion of the
Bogolubov-Born-Green-Kirkwood-Yvon (BBGKY) hierarchy of equations
for the reduced density operators \cite{Bonitz}. Let us note that
this approach was applied when we studied transient spectra of
resonance fluorescence \cite{Roerich2003}.

As the value of the coupling constant is usually small it is sufficient to use just the Born approximation
to study the radiative relaxation processes. In the framework of this approach we write
 \be \label{Smain3}
    \begin{array}{c}
     \dsp{i\frac{d\ra}{dt}}-\sum_{\kl}\Tr_f[\Vafo,\rft\ra]= \sum_{\kl}\Tr_f[\Vafo,\gaf],\\
     \dsp{i\frac{d\rft}{dt}}-\Tr_a[\Vafo,\rft\ra]\Spa=\Tr_a[\Vafo,\gaf],\\
     \dsp{i\frac{d\gaf}{dt}}=[\Vafo,\ra\rft],\\
  \end{array}
\ee where $\ra,\rft$ are the atomic and field single particle density operators respectively, while $\gaf$
is the single atom - single photon correlation operator. $\Tr_a,\Tr_f$ denote tracing operation over field
and atomic variables.

Operator $\Tr_a[\Vafo,\rft\ra]$ describes polarization of the medium induced by the external field and
$\sum_{\kl}\Tr_f[\Vafo,\rft\ra]$ introduces the mean field contribution. Since our focus is to study the
relaxation processes only we further disregard the contribution of these terms.

Finding a formal solution for $\gaf$ we substitute it to the
 equations for the atomic and field density operators:
  \be
    \label{Smain4}
       \begin{array}{c}
        \dsp{i\frac{d\ra}{dt}}=\sum_{\kl}\Tr_f[\Vafo(t),\int_0^t[\Vafo(\tau),\ra(\tau)\rft(\tau)]d\tau], \\
        \dsp{i\frac{d\rft}{dt}}=\Tr_a[\Vafo(t),\int_0^t[\Vafo(\tau),\ra(\tau)\rft(\tau)]d\tau].\\
     \end{array}
 \ee
 The right-hand-side terms in the equation for the atomic density
 operator is the operator which describes both stimulated and
 spontaneous transitions while the right-hand-side terms in the
 equation for the field density operator describes the variation of
 the number of photons due to such atomic transitions.

\section{\label{Spon}Spontaneous transitions}
To consider properties of spontaneous transitions one needs to
substitute $\rft(\tau)\equiv |0_{\bkl}\rangle\langle 0_{\bkl}|$ to
the right-hand side of the equation for the density operator. It
is clear that the terms proportional to
$\am\ap\rft,\rft\am\ap,\ap\rft\am$ remain while the others vanish.
Taking the latter into account it is now possible to follow the
well known Wigner-Weisskopf theory to transform to
\begin{widetext}
  \be
    \label{Smain5}
       \begin{array}{c}
   \dsp{i\frac{d\ra}{dt}=-i\sum_{j_1,j_2=b,c}\sum_{M_{j_1}M_{j_2}} }  \Bigl(\Gamma_{j_1
   j_2}(M_{j_1},M_{j_2})e^{i(\omega_{j_1 d}-\omega_{j_2 d})t}|j_1M_{j_1}\rangle\langle M_{j_2}
   j_2|\ra + \\
   \dsp{-\sum_{M_{d_1}M_{d_2}} \Gamma_{j_1j_2}(M_{j_1} M_{d_1} ,M_{j_2} M_{d_2} )}e^{i(\omega_{j_1 d}-\omega_{j_2 d})t}|d M_{d_1}\rangle\langle M_{j_1}
   j_1|\ra|j_2 M_{j_2}\rangle\langle M_{d_2} d| +H.c. \Bigr), \\
   \end{array}
 \ee
  \be
    \label{Smain6}
       \begin{array}{c}
   \dsp{ \Gamma_{j_1 j_2}(M_{j_1},M_{j_2})= \int \sum_{M_{d} \sigma \sigma^{'}}}\sum_{\lambda=\pm1}
  g_{j_1 d}^{{\bf k}\lambda}(M_{j_1} M_d) g_{d j_2}^{{\bf k}\lambda}(M_d M_{j_2})\delta(\omega_{j_2 d}-\wk)
  \dsp{\frac{W d {\bf k}}{(2\pi)^3}}
   \dsp{=\sum_{M_d \sigma,\sigma^{'}} \Qind },\\
    \dsp{\Gamma_{j_1j_2}(M_{j_1} M_{d_1} ,M_{j_2} M_{d_2} )= \int \sum_{\lambda=\pm1} g_{j_1 d}^{{\bf k}\lambda}(M_{j_1} M_{d_1})
    g_{d j_2}^{{\bf k}\lambda}(M_{d_2} M_{j_2})\delta(\omega_{j_2 d}-\wk)
  \frac{W d {\bf k}}{(2\pi)^3}}
  \dsp{=\Qind }\\
  \dsp{\Qind=S_{J_{j_1} J_{j_2}} C_{J_d M_d 1\sigma}^{J_{j_1} M_{j_1}} C_{J_d M_d 1\sigma^{'}}^{J_{j_2} M_{j_2}}}K^R(\sigma,\sigma^{'}),\\
  \dsp{ S_{j_1 j_2}=\frac{2\parallel \mu_{j_1 d}\parallel\parallel \mu_{j_2 d}\parallel \omega_{j_2 d}^3}{\hbar c^3\sqrt{(2J_{j_1}+1)(2J_{j_2}+1)}},\;\;\;\;
   K^R(\sigma,\sigma^{'})=\int\sum_{\lambda=\pm1}\Bigl( D_{\lambda\sigma^{'}}^1\Bigr)^{*} D_{\lambda\sigma}^1
   \frac{dO}{8\pi^2}},
    \end{array}
 \ee
 \end{widetext}
where $dO$ is the spatial angle element. Now by using the
$\ra\to e^{i\Ha t}\ra e^{-i\Ha t}$ transformation we change the equation \rf{Smain5} to the from with the constant coefficients:
  \be
    \label{Smain7}
       \begin{array}{c}
   \dsp{i\frac{d\ra}{dt}-[\Ha,\ra]-\Gamma^R[\ra]=0} \\
  \end{array}
 \ee
where $\Gamma^R[\ra]$ is the relaxation operator for a degenerate
system of $V$-type.
 \begin{widetext}
  \be
    \label{Smain8}
       \begin{array}{c}
   \dsp{ \Gamma^R[\ra]=-i\sum_{j_1,j_2=b,c}\sum_{M_{j_1}M_{j_2}}\Bigl(\Gamma_{j_1
   j_2}(M_{j_1},M_{j_2})|j_1M_{j_1}\rangle\langle M_{j_2}
   j_2|\ra  +\ra \Gamma_{j_1j_2}( M_{j_1},M_{j_2} )} \dsp{ |j_2M_{j_2}\rangle\langle M_{j_1}
   j_1|}\\
   \dsp{-\sum_{M_{d_1}M_{d_2}} \Gamma_{j_1j_2}(M_{j_1} M_{d_1} ,M_{j_2} M_{d_2} )}\bigl(|d M_{d_1}\rangle\langle M_{j_1}
   j_1|\ra|j_2 M_{j_2}\rangle\langle M_{d_2} d|+|d M_{d_1}\rangle\langle M_{j_2}
   j_2|\ra|j_1 M_{j_1}\rangle\langle M_{d_2} d|\bigr) \Bigr). \\
   \end{array}
 \ee
\end{widetext}
In Eq.(\ref{Smain8}) the first two terms inside brackets describe radiative relaxation of the density
operator and transitions to the lower
 levels while the third term describes populating of the lower levels. Eqs. \rf{Smain6} and \rf{Smain8} determine the
  relaxation processes completely for the model under consideration. In \cite{Roer2000.2} we showed that the operator
   in charge of radiative depopulation is of diagonal structure for most of real systems. It follows directly from
   orthogonality of the Wigner functions \cite{Landau3} that:
\begin{equation}
 \label{Ortog}
   \int \Bigl(D_{\lambda \sigma_1}^{J_1}\Bigr)^{*} D_{\lambda \sigma_2}^{J_2}\frac{dO}{4\pi}=\frac{1}{2J+1}
   \delta_{J_1 J_2}\delta_{\sigma_1 \sigma_2},
\end{equation}
allowing for the property \cite{Varshal}:
\begin{equation}
\label{SummK}
 \sum_{M_d} C_{J_d M_d 1\sigma}^{J_{j_1} M} C_{J_d M_d 1\sigma}^{J_{j_2} M}=\delta_{J_{j_1}J_{j_2}}.
\end{equation}
From Eqs. (\ref{Ortog})(\ref{SummK}) it follows  that $\Gamma_{j_1 j_2}(M_{j_1} M_{j_2})$ is different
from $0$ only if $j_1=j_2$ and $M_{j_1}=M_{j_2}$.

It must be noted that this property of the Wigner functions \rf{Ortog} and the properties of the dipole
transition moments stated in \cite{quench-zhu}, \cite{qbeat-patnaik},
\cite{qbeat-patnaik1},\cite{quench-berman},\cite{aniso-gsa} are indeed the expressions for the same common
property which reflects the fact that interference is only possible for the fields with the same
projections of the magnetic moment in the atomic coordinates. However, as it was demonstrated earlier it
is insufficient for the dipole transition moments to be parallel to ensure $\Gamma_{j_1j_2}(
M_{j_1},M_{j_2} )$ being different from zero.

The properties of the coefficients which describe the populating processes for the lower levels are
somewhat similar to those discussed above, i.e., the consequence of the orthogonal property of Wigner's
functions \rf{Ortog}. Factors $\Gamma_{j_1j_2}(M_{j_1} M_{d_1} ,M_{j_2} M_{d_2} )$ are different from zero
only if $M_{j_1}-M_{d_1}=M_{j_2}-M_{d_2}$. However, it must be noted that in a general case  the
relaxation operator describing the process of population of the lower levels is not of the diagonal
structure. Off-diagonal terms describe so called polarization transport at spontaneous transitions. This
process was analysed in \cite{Rautian}.

Let us note that our analysis presumes implicitly that the
influence of the environment around the atom upon the rate and the
properties of the spontaneous radiation is rather weak. However, a
different situation could be observed in a medium with modified
density of states of electromagnetic modes. In this case the
off-diagonal terms in the relaxation operator could be different
from zero. We shall mark out the following cases:

1. The atoms are placed in a high-Q cavity which changes the density of states of electromagnetic modes.
In this case the cavity's axis determines the designated direction (quantization axis). The density of
states for photons of different polarization can vary significantly so the interference cancellation in
(\ref{SummK}) is broken. Naturally, one has to take account for the Parcell effect being the change of
conventional spontaneous decay rates $\Gamma_{j_1j_1}$ in cavities.

As an example we shall adduce the values of $K^R(\sigma,\sigma^{'})$ for an atom placed between two thin
plates with the reflection factor $r$ and the inter-plate distance $d$ satisfying $kd\ll 1$. The normal
line to the plate's plane is chosen to be the quantization axis. It is easy to see that due to the chosen
symmetry $K^R(\sigma,\sigma^{'})=0$ if $\sigma\ne\sigma^{'}$. Besides, $K^R(1,1)=K^R(-1,-1)\sim
\gamma_\perp$, $K^R(0,0)\sim \gamma_\parallel$, where $\gamma_\perp,\gamma_\parallel$ are the
probabilities of spontaneous transitions for dipoles oriented either normally or parallel to the plate's
plane respectively. We will use the results of \cite{LoudonSp} where it had been demonstrated for such a
configuration that
 \be
   \label{SpontConst}
    \gamma_\perp=\gamma\frac{1+|r|}{1-|r|},\;\;\;\;\gamma_\parallel=\gamma\frac{1-|r|}{1+|r|}
 \ee
where $\gamma$ is the spontaneous decay rate in vacuum.
  \begin{equation}
  \label{SpontConst1}
   \begin{array}{c}
     \dsp{K^R(1,1)=K^R(-1,-1)=K^R_{vac}(1,1)\frac{1-|r|}{1+|r|} }, \\
     \dsp{ K^R(0,0)=K^R_{vac}(0,0)\frac{1+|r|}{1-|r|}. }\\
    \end{array}
  \end{equation}

Thus, Eq.(\ref{SummK}) cannot be realized and the off-diagonal terms in the relaxation operator are
different from zero. Moreover, for $r \approx 1$ the dipole transitions oriented parallel to the plate's
plane are cancelled completely. As a consequence, the only possible transitions are those which are
oriented normally to  the plates. Thus, all possible transitions have parallel dipole moments and the
magnitude of the off-diagonal term is maximum in this case. Namely, if, similarly  to \cite{quench-zhu},
\cite{qbeat-patnaik}, \cite{qbeat-patnaik1},\cite{quench-berman}, $p$ is defined as $\Gamma_{j_1
j_2}(M,M)=p\sqrt{\Gamma_{j_1 j_1}(M,M)\Gamma_{j_2 j_2}(M,M)}$ and used as the factor determining the
quantum interference effects, in this case $|p|\approx 1$.

2. The atoms are placed in a photonic crystal. In this case,  one can as well see a change in the density
of states of the electromagnetic field with respect to polarization as it was discussed in the previous
paragraph. However, there is also frequency variation of the density of the field states \cite{John}
giving rise to the frequency gaps. A transition within this gap is completely quenched \cite{Yablonovich}
as well as the corresponding off-diagonal terms should turn to zero. If it is assumed that the transition
frequencies are the functions of the magnetic moment projection, e.g., in a magnetic field, one can set
the frequency gap to include one of the transitions with a definite $\sigma$ which would not meet the
condition for Eq.(\ref{SummK}). In its turn that would make the off-diagonal terms of the relaxation
operator different from zeros.

Let us note that for a photonic crystal there should be a strong
condition for $\Gamma_{j_1 j_2}\ne\Gamma_{j_2 j_1}$ to be true. As
one can see from Eq.(\ref{Smain6}) this effect could be observed
for a free atom as well, however, being utterly insignificant in
value and of the order of $O(3(\omega_{j_1 d}-\omega_{j_2
d})/\omega_{j_1 d})$. This is explained by the fact that in a free
space the density of the field modes is proportional to
$N(\wk)=2\wk^2/ c^3$ and varies weakly for different transitions
as long as the frequency difference is much smaller than the
transition frequency. For an isotropic photonic crystal the mode
density near the upper bound of the frequency gap $\omega_e$ has
the form \cite{John}:
 $$
  \begin{array}{c}
    \wk=\omega_e+A(k-k_0)^2, \\
    \dsp{ N(\wk)=\sqrt{\frac{\wk-\omega_e}{A^3}}\Theta(\wk-\omega_e)}\\
   \end{array}
 $$
which should result in strong frequency dependence of $\Gamma_{j_1
j_2}$.

Thus, if the density of states of the electromagnetic field is
anisotropic and the probabilities of photon emission with
different polarization are unequal the off-diagonal terms in the
relaxation operator are different from zero. A similar situation
takes place for the operator in charge of stimulated transitions
when an atom is considered under action of the field of thermal
photons. We shall discuss this matter in detail in the following
section.

\section{\label{Stim}Stimulated transitions}
We assume that a single-photon density operator is in its excited state which  characterizes some
incoherent field (not necessarily meeting the Planck distribution) and has the diagonal structure:
 \be
  \rft=\sum_{n_{\bkl}}\rft(n_{\bkl})|n_{\bkl}\rangle\langle
  n_{\bkl}|.
 \ee
In this case one can define the average number of photons in a mode as
 \be
  N_{\bkl}=\sum_{n_{\bkl}}\langle
  n_{\bkl}|\ap\am\rft|n_{\bkl}\rangle=\sum_{n_{\bkl}}n_{\bkl}\rft(n_{\bkl}).
 \ee
This number can be a function of vector $\bk$ and have a
non-trivial angular and polarization distribution.

We now will examine the properties of the transition operator \rf{Smain4} taking into account the
properties of the field density operator stipulated above. Unlike the situation considered in the previous
section, the terms different from zero are not only those proportional to
$\am\ap\rft,\rft\am\ap,\ap\rft\am$, but also the ones proportional to $\ap\am\rft,\rft\ap\am,\am\rft\ap$.
Following the steps of Wigner-Weisskopf theory and the steps taken in the previous section we get the
equation for the atomic density operator in the form:
   \be
    \label{Smain9}
       \begin{array}{c}
   \dsp{i\frac{d\ra}{dt}-[\Ha,\ra]-\Gamma^S[\ra]-\Gamma^R[\ra]=0}, \\
  \end{array}
  \ee
where operator $\Gamma^S[\ra]$ describes transitions in a
degenerate $V$-type system
  \begin{widetext}
  \be
    \label{Smain10}
       \begin{array}{c}
   \dsp{ \Gamma^S[\ra]=-i\sum_{j_1,j_2=b,c}\sum_{M_{j_1}M_{j_2}}\Bigl(\Gamma_{j_1
   j_2}^S(M_{j_1},M_{j_2})|j_1M_{j_1}\rangle\langle M_{j_2}
   j_2|\ra  +\ra \Gamma_{j_1j_2}^S( M_{j_1},M_{j_2} )}  \dsp{  |j_2M_{j_2}\rangle\langle M_{j_1}
   j_1|} \\
   \dsp{-\sum_{M_{d_1}M_{d_2}} \Gamma_{j_1j_2}^S(M_{j_1} M_{d_1} ,M_{j_2} M_{d_2} )}
   \bigl(|d M_{d_1}\rangle\langle M_{j_1}j_1|\ra|j_2 M_{j_2}\rangle\langle M_{d_2} d|
   +|d M_{d_2}\rangle\langle M_{j_2} j_2|\ra|j_1 M_{j_1}\rangle\langle M_{d_1} d|\bigr) \Bigr)- \\
  \dsp{-i\sum_{M_{d_1}M_{d_2}} \Bigl(\Gamma_{d d}^S(M_{d_1},M_{d_2})|d M_{d_1}\rangle\langle M_{d_2} d|\ra
   +\ra \Gamma_{dd}^S( M_{d_1},M_{d_2}) |d M_{d_2}\rangle\langle M_{d_1} d|-}\\
    \dsp{-\sum_{j_1 j_2=b,c}\sum_{M_{j_1}M_{j_2}} \Gamma_{dd}^S(M_{d_1} M_{j_1} ,M_{d_2} M_{j_2} )}\bigl(|j_1 M_{j_1}\rangle\langle M_{d_1}
   d|\ra|d M_{d_2}\rangle\langle M_{j_2} j_2|+|j_2 M_{j_2}\rangle\langle M_{d_2}
   d|\ra|d M_{d_1}\rangle\langle M_{j_1} j_1|\bigr)\Bigr), \\
    \end{array}
 \ee
\be
    \label{Smain11}
       \begin{array}{c}
   \dsp{ \Gamma_{j_1 j_2}^S(M_{j_1},M_{j_2})= \int \sum_{M_{d} \sigma \sigma^{'}}}\sum_{\lambda=\pm1}
  g_{j_1 d}^{{\bf k}\lambda}(M_{j_1} M_d) g_{d j_2}^{{\bf k}\lambda}(M_d M_{j_2})\delta(\omega_{j_2 d}-\wk)
  \dsp{\frac{W d {\bf k}}{(2\pi)^3}}
   \dsp{=\sum_{M_d \sigma,\sigma^{'}} \Bind}   ,\\
    \dsp{\Gamma_{j_1j_2}^S(M_{j_1} M_{d_1} ,M_{j_2} M_{d_2} )= \int \sum_{\lambda=\pm1} g_{j_1 d}^{{\bf k}\lambda}(M_{j_1} M_{d_1}) g_{d j_2}^{{\bf k}\lambda}(M_{d_2} M_{j_2})\delta(\omega_{j_2 d}-\wk)
  \frac{W d {\bf k}}{(2\pi)^3}}
  \dsp{=\Bind },\\
      \end{array}
 \ee
 $$
 \begin{array}{c}
 \dsp{ \Gamma_{d d}^S(M_{d_1},M_{d_2})= \int \sum_{j=b,c} \sum_{M_{j} \sigma \sigma^{'}}}\sum_{\lambda=\pm1}
  g_{d j}^{{\bf k}\lambda}(M_d M_j) g_{jd}^{{\bf k}\lambda}(M_j M_d)\delta(\omega_{jd}-\wk)
  \dsp{\frac{W d {\bf k}}{(2\pi)^3}}
  \dsp{=\sum_{j=b,c} \sum_{M_j \sigma,\sigma^{'}} B^{J_j M_j J_j M_j}_{J_d M_d \sigma \sigma^{'}} },\\
  \dsp{\Gamma_{dd}^S(M_{d_1} M_{j_1} ,M_{d_2} M_{j_2} )= \int \sum_{\lambda=\pm1}g_{d j_1}^{{\bf k}\lambda}(M_{d_1} M_{j_1}) g_{j_2 d}^{{\bf k}\lambda}(M_{j_2} M_{d_2})\delta(\omega_{j_2d}-\wk)
   \frac{W d {\bf k}}{(2\pi)^3}
   =\Gamma_{j_1j_2}^S(M_{j_1} M_{d_1} ,M_{j_2} M_{d_2} )},\\
  \dsp{ \Bind=S_{j_1 j_2}C_{J_d M_d 1\sigma}^{J_{j_1} M_{j_1}} C_{J_d M_d 1\sigma^{'}}^{J_{j_2}
    M_{j_2}}K^S(\sigma,\sigma^{'})},\\
   \dsp{K^S(\sigma,\sigma^{'})=\int\sum_{\lambda=\pm1}N_{\bkl}(\wk=\omega_{j_2d})\Bigl( D_{\lambda\sigma^{'}}^1\Bigr)^{*} D_{\lambda\sigma}^1
   \frac{dO}{8\pi^2}}. \\
    \end{array}
 $$
\end{widetext}

In Eq.\rf{Smain10} the terms in the first parentheses describe stimulated transitions emitting a photon,
while the terms in the second parentheses describe stimulated transitions with absorption of a photon. In
case the photonic field is homogeneous and the average number of photons in each mode does not depend on
the wave vector direction the properties of $\Gamma_{j_1 j_2}^S(M_{j_1},M_{j_2}),\Gamma_{j_1j_2}^S(M_{j_1}
M_{d_1} ,M_{j_2} M_{d_2}); j_1,j_2=b,c,d$ are absolutely the same as those of the relaxation operator
components $\Gamma^R[\ra]$. This is conditioned by the fact that we can exclude $N_{\bkl}$ from the
integration and use the orthogonality properties of the Wigner functions \rf{Ortog} and of the
Glebsch-Gordan coefficients \rf{SummK}. In this case, similar to the relaxation operator for
$\Gamma_{j_1j_2}^S(M_{j_1},M_{j_2}),\Gamma_{d d}^S(M_{d_1},M_{d_2})$ the terms different from zero are
those being diagonal, i.e., $j_1=j_2,M_{j_1}=M_{j_2},M_{d_1}=M_{d_2}$. Coefficients
$\Gamma_{j_1j_2}^S(M_{j_1} M_{d_1} ,M_{j_2} M_{d_2}), \Gamma_{dd}(M_{d_1} M_{j_1} ,M_{d_2} M_{j_2} )$ are
non-zero only for $M_{j_1}-M_{d_1}=M_{j_2}-M_{d_2}$.

In case the photonic field is anisotropic the orthogonality property \rf{Ortog} is not applicable so the
off-diagonal terms can be non-zero.  It makes the relaxation operator complete and rather complicated. As
an example, we shall consider and compare the structure of the relaxation terms $\Gamma_{j_1
j_2}^S(M_{j_1},M_{j_2})$ for $D$-line ($P_{3/2}\to S_{1/2},P_{3/2}\to P_{1/2}$) of alkaline metal vapor
when the photonic field has relatively simple axisymmetric along the (z) axis structure.
 \be
    \label{field}
    N_{\bkl}(\wk=\omega_{j_2d})=N\cos^2(\theta).
 \ee
The integration over the spatial angle has the form:
 \be
    \label{integral}
     \begin{array}{c}
    \dsp{K^S(\sigma,\sigma^{'})=\sum_{\lambda=\pm1}\int_0^{2\pi} e^{i(\sigma-\sigma^{'})\phi}d\phi}\times \\
    \dsp{ \times \int_0^{\pi} N_{\bkl}(\wk=\omega_{j_2d}) s_{\lambda\sigma^{'}}^1
    s_{\lambda\sigma}^1\frac{d\theta}{8\pi^2}}. \\
      \end{array}
 \ee
To perform further calculations we need the explicit expressions
for $s_{m^{'} m}^1(\beta)$ which determine the transformations of
the wave functions when the coordinate system is rotated around
the ``knot''  line \cite{Landau3}:
 \be
  \label{Dsmall}
  \begin{array}{c}
  s_{1 0}^1=s_{0 -1}^1=-s_{0 1}^1=-s_{-10}^1=
\dsp{\frac{1}{\sqrt{2}}\sin{\beta}}, \\
  s_{1 1}^1=s_{-1 -1}^1=\dsp{\frac{1}{2}(1+\cos{\beta}}),\\
  s_{1 -1}^1=s_{-1 1}^1=\dsp{\frac{1}{2}(1-\cos{\beta}}).
  \end{array}
\ee
Both in case of the homogeneous field and the field with the
axisymmetric along the (z) axis distribution
$K(\sigma,\sigma^{'})=0$ at $\sigma\ne\sigma^{'}$ which is
conditioned by integration over angle $\phi$. Consequently, the
constants different from zero are those which are of the form
$\Gamma_{j_1 j_2}^S(M,M)$.

Now we list the values of $\Gamma_{j_1 j_2}^S(M_,M)$ expressed
through $K^S(\sigma, \sigma)$ (see Figs. 2,3):
 \be
 \label{GammFree}
  \begin{array}{c}
  \dsp{\Gamma_{bb}^S\left( \frac{3}{2},\frac{3}{2} \right)=S_{bb}K^S(1,1)},\\
  \dsp{\Gamma_{bb}^S\left(-\frac{3}{2},-\frac{3}{2}\right)=S_{bb}K^S(-1,-1)}, \\
  \dsp{\Gamma_{bb}^S\left(\frac{1}{2},\frac{1}{2}\right)=S_{bb}\frac{K^S(1,1)+2K^S(0,0)}{3}},\\
  \dsp{\Gamma_{bb}^S\left(-\frac{1}{2},-\frac{1}{2}\right)=S_{bb}\frac{K^S(-1,-1)+2K^S(0,0)}{3}}, \\
  \dsp{\Gamma_{cc}^S\left(\frac{1}{2},\frac{1}{2}\right)=S_{cc}\frac{2K^S(1,1)+K^S(0,0)}{3}},\\
  \dsp{\Gamma_{cc}^S\left(-\frac{1}{2},-\frac{1}{2}\right)=S_{cc}\frac{2K^S(-1,-1)+K^S(0,0)}{3}}, \\
   \dsp{\Gamma_{bc}^S\left(\frac{1}{2},\frac{1}{2}\right)=S_{bc}\frac{\sqrt{2}}{3}(K^S(0,0)-K^S(1,1)),} \\
  \dsp{\Gamma_{bc}^S\left(-\frac{1}{2},-\frac{1}{2}\right)=S_{bc}\frac{\sqrt{2}}{3}(K^S(-1,-1)-K^S(0,0)),}\\
  \dsp{\Gamma_{j_1 j_2}^S\left(\pm\frac{1}{2},\pm\frac{1}{2}\right)=\Gamma_{j_2 j_1}^S\left(\pm\frac{1}{2},\pm\frac{1}{2}\right)}.\\
  \end{array}
\ee
\begin{figure}
    \includegraphics[scale=0.85]{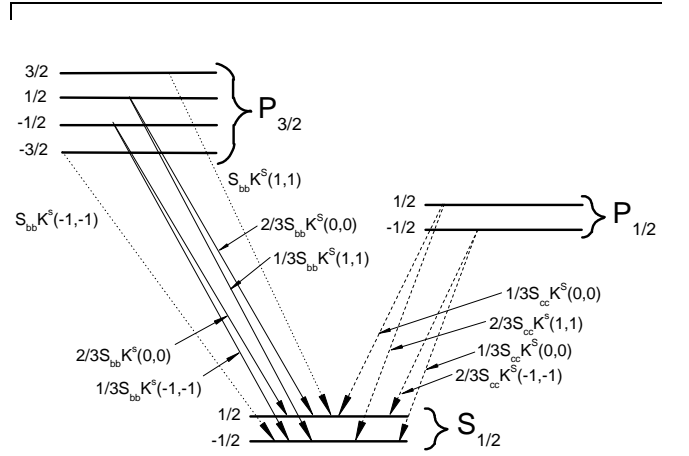}
   \caption{Transitions for the $D$-line of alkali metals. }
  \label{fig:2}
  \end{figure}
\begin{figure}
    \includegraphics[scale=0.85]{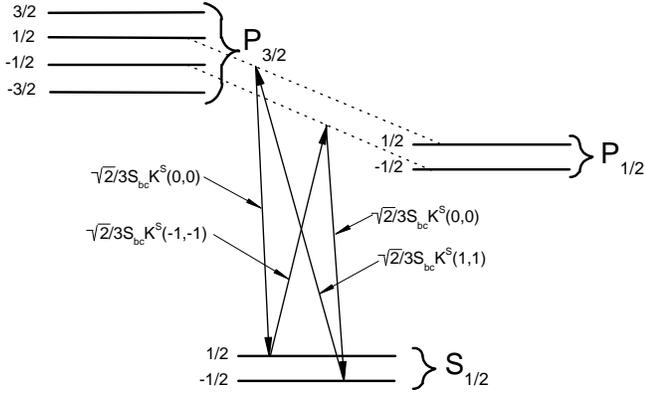}
   \caption{Additional interference transitions for the $D$-line of alkali metals. }
  \label{fig:3}
  \end{figure}
For $K^S(\sigma,\sigma)$ and homogeneous field we have:
 \be
   \label{Const1}
    \begin{array}{c}
       \dsp{K^S(0,0)=N
       \int_{0}^{\pi}\sin^2(\theta)\frac{\sin{\theta}}{2}d\theta=N\frac{2}{3},
     }\\
     \dsp{K^S(1,1)=K^S(-1,-1)=} \\ \dsp{
     =N\int_{0}^{\pi}\frac{(1+\cos^2(\theta))}{2}\frac{\sin{\theta}}{2}d\theta=N\frac{2}{3}.
     }\\
     \end{array}
 \ee
Let us mark out the well known fact that:
\begin{equation}
 \label{Soot}
 \displaystyle{\frac{||\mu_{bd}||}{\sqrt{2J_b+1}}=\frac{||\mu_{cd}||}{\sqrt{2J_c+1}}},
\end{equation}
which makes all the constants $S_{j_1 j_2}=S$ equal from which it follows that with  account for
\rf{Const1} and $\Gamma_{j j}^S(M,M)$ the values for magnetic sublevels also become equal \footnote{From
which it directly follows (see \rf{Smain6}), that the factors to the relaxation terms for magnetic
sublevels $\Gamma_{j j}^R(M,M)$ in atoms of alkali metals are equal.}
 \be
   \dsp{\Gamma_{j j}^S(M,M)=S N \frac{2}{3}.}
 \ee
Consequently, just as we expected, the off-diagonal terms $\Gamma_{b c}^S (M,M)$ appeared to be strictly
equal to zero.

For the field distribution \rf{field} factors
$K^S(\sigma,\sigma^{'})$ have the form:
 \be
   \label{Const2}
    \begin{array}{c}
       \dsp{K^S(0,0)=N
       \int_{0}^{\pi}\sin^3(\theta)\frac{\cos^2{\theta}}{2}d\theta=N\frac{4}{15},
     }\\
     \dsp{K^S(1,1)=K^S(-1,-1)=} \\ \dsp{
     =N\int_{0}^{\pi}\frac{(1+\cos^2(\theta))\cos^2(\theta)}{2}\frac{\sin{\theta}}{2}d\beta=N\frac{4}{75}.
          }\\
     \end{array}
 \ee
Taking into account \rf{GammFree}, \rf{Soot} factors $\Gamma_{j_1
j_2}^S(M,M)$ can be written as
 \be
 \label{GammFree1}
  \begin{array}{c}
  \dsp{\Gamma_{bb}^S\left(\frac{3}{2},\frac{3}{2}\right)=\Gamma_{bb}^S\left(-\frac{3}{2},-\frac{3}{2}\right)={12}/{225} N S },\\
  \dsp{\Gamma_{bb}^S\left(\frac{1}{2},\frac{1}{2}\right)=\Gamma_{bb}^S\left(-\frac{1}{2},-\frac{1}{2}\right)={44}/{225} N S},\\

  \dsp{\Gamma_{cc}^S\left(\frac{1}{2},\frac{1}{2}\right)=\Gamma_{cc}^S\left(-\frac{1}{2},-\frac{1}{2}\right)={28}/{225} N S},\\
  \dsp{\Gamma_{bc}^S\left(\frac{1}{2},\frac{1}{2}\right)=-\Gamma_{bc}^S\left(-\frac{1}{2},-\frac{1}{2}\right)={16\sqrt{2}}/{225} N S .} \\
  \end{array}
 \ee
Thus, if an atom interacts with an anisotropic and incoherent field the factors $\Gamma_{j_1
j_2}^S(M_{j_1},M_{j_2})$, which determine induced transitions emitting photons, change considerably both
in the absolute value and the structure, or in other words, some additional off-diagonal terms are
produced. It must be noted that for the configuration chosen the value of the off-diagonal term becomes
quite substantial. Namely, if by analogy with  previous section we define the factor $p$ as $\Gamma_{j_1
j_2}(M,M)=p\sqrt{\Gamma_{j_1 j_1}(M,M)\Gamma_{j_2 j_2}(M,M)}$  and used it as the gauge for quantum
interference, then for transitions $P_{3/2}(\pm1/2)\to S_{1/2}(\pm1/2),P_{1/2}(\pm1/2)\to S_{1/2}(\pm1/2)$
this relation would take the form:
  \be
    \begin{array}{c}
       \dsp{\Gamma_{bc}^S\left(\pm\frac{1}{2},\pm\frac{1}{2}\right)\approx } \\
   \dsp{    \approx \pm0.64
    \sqrt{\Gamma_{bb}^S\left(\pm\frac{1}{2},\pm\frac{1}{2}\right)\Gamma_{cc}^S\left(\pm\frac{1}{2},\pm\frac{1}{2}\right)}}.\\
    \end{array}
  \ee

\section{Conclusions}
Now in this section we summarize the main results of this work. We used the approach based on the cluster
expansion of the BBGKY hierarchy for the reduced density operators and the Wigner-Weisskopf approximation
to derive explicitly the relaxation terms for $V$-type degenerate systems with an arbitrary total angular
momentum of each level. Our main focus was to study the effect of quantum interference on the structure
and properties of the relaxation terms. The radiation emission term has been shown to be diagonal for most
real systems. The condition for the dipole transition moments to be parallel, discussed in
\cite{quench-zhu}, \cite{qbeat-patnaik}, \cite{qbeat-patnaik1}, and \cite{quench-berman}, appears to be
insufficient to give rise to quantum interference. We show that some additional conditions are necessary,
one of which is anisotropy of the density of states of the electromagnetic field or its strong frequency
dependence. This fact also follows from the analysis of the relaxation operator carried out in Appendix
with account for the hyperfine splitting of atomic levels. This conditions can be well realised for a
high-Q cavity or a photonic crystal. In this case the interference terms can reach their maximum
theoretically predicted magnitude.

For stimulated transitions the condition for the quantum
interference to occur can be substantially weakened. This case was
considered in Sec. \rf{Stim} where we studied the effect of
quantum interference upon the structure of the terms describing
stimulated transitions in the external photonic field. It has been shown that for such field the additional
off-diagonal terms are likely to appear. This case was thoroughly
analysed for the $D$-line in the spectrum of alkaline metal vapour
as well as for the atom interacting with an axisymmetric field.

The stimulated transition operator has been derived and shown to
have the off-diagonal terms as high as $64\%$ of the maximum of
theoretically predicted value for the chosen field configuration.

\section{Acknowledgement}
The authors are grateful to Prof. A.N. Starostin for fruitful discussions and valuable remarks. This work
was supported by the Russian Foundation for Basic Research 02-02-17153 , as well as by grants MK
1565.2003.02, and NSh 1257.2003.02 of the President of the Russian Federation.

\section{\appendixname. Relaxation operator, hyperfine splitting.}

In this section we will derive explicitly the relaxation operator
for a doublet $V$-type atom with account for the hyperfine
splitting. Under this approximation the Hamiltonian (in $1/\hbar$
units) can be written a sum:
\begin{equation}
 \label{GamiltonF}
  \hat{H}=\Ha+\sum_{\kl}\Hf+\sum_{\kl}\Vaf.
\end{equation}
In (\ref{GamiltonF}) $\Ha$ describes an unperturbed
system:
\begin{equation}
 \label{GamH0F}
  \Ha=\sum_{j=b,c,d}\sum_{F_j}\sum_{M_F}\omega_{j}(F_j)|j F_j M_{F_j}\rangle\langle M_{F_j} F_j j|,
\end{equation}
where $\hbar\omega_{j}(F_j)$ is the atomic state energy with the total angular momentum $J_j$, total
momentum $F_j$ and its projection $M_F^j$ to the quantization axis (z), $|j,F_j,M_{F}\rangle$ is the
corresponding vector.

The next term in \rf{GamiltonF} is nothing but the one used in
Sec. \ref{bas} and describes the Hamiltonian of the quantized
field:
\begin{equation}
 \label{GamHfF}
  \hat{H_f}=\wk\ap\am,
\end{equation}
The atom-field interaction is described by
\begin{equation}
 \label{GamVqF}
 \begin{array}{c}
  \dsp{\Vaf=i\sum_{j=b,c}\sum_{F_j F_d}\sum_{M_{F_j} M_{F_d}}} \\
  \dsp{  g^{{\bf k}\lambda}_{jd}(F_j M_{F_j};F_d M_{F_d})\am|j F_j M_{F_j}\rangle \langle M_{F_d} F_d d| }+ H.c \\
 \end{array}
\end{equation}
where
 \be
   \label{constF}
   \begin{array}{c}
   \dsp{g^{{\bf k}\lambda}_{jd}(F_j M_{F_j}; F_d M_{F_d})=} \\ \dsp{ =\sqrt{\frac{2\pi\wk}{\hbar
   W}}\left(\vec{\mu}_{jd}(F_j M_{F_j}; F_d M_{F_d}),
  \epsilon(\bkl)\right) },
  \end{array}
 \ee
describes the coupling constant with $\vec{\mu}_{jd}(F_j M_{F_j};
F_d M_{F_d})$ being the transition dipole moment. Using the Wigner
Ekkart theorem \cite{Varshal} and taking into account the fact
that the transition dipole moment commutes with the nucleus spin
\rf{const} the coupling constant can be written as
\cite{Sobelman}:
 \begin{widetext}
 \be
  \label{constF1}
    \begin{array}{c}
 \dsp{ g^{{\bf k}\lambda}_{jd}(F_j M_{F_j}; F_d M_{F_d})=(-1)^{I+1-F_j-F_d} }
   \dsp{\sqrt{(2F_d+1)} W(J_jF_jJ_d F_d;I1)}
  \dsp{ C^{F_j M_{F_j}}_{ F_d M_{F_d} 1\sigma} ||\mu_{jd}||
   \sqrt{\frac{2\pi\wk}{\hbar W}}D_{\lambda\sigma}^1},
   \end{array}
 \ee
where $W(J_jF_jJ_d F_d;I1)$ are the Racah coefficients and $I$ is
the nucleus spin. Now leaving out the transformations similar to
those described in Sec. \ref{Spon} we straightforwardly write the
relaxation operator in the following form:
 \be
    \label{Smain8F}
       \begin{array}{c}
   \dsp{ \Gamma^R[\ra]=-i\sum_{j_1,j_2=b,c}\sum_{F_1 F_2}\sum_{M_1 M_2}\Bigl(\Gamma_{j_1
   j_2}(F_1 M_1,F_2 M_2)|j_1 F_1 M_1\rangle\langle M_2 F_2
   j_2|\ra   } \\
   \dsp{-\sum_{M_{d_1}M_{d_2}} \Gamma_{j_1j_2}(F_1 M_1, F_1^d M_1^d ;F_2 M_2, F_2^d M_2^d )}|d F_2^d M_2^d\rangle\langle
   M_2 F_2 j_2|\ra|j_1 F_1 M_1\rangle\langle M_1^d F_1^d d| + H.c.\Bigr), \\
   \end{array}
 \ee
where for the sake of simplicity we introduced the following
notations: $F_{j_1}\to F_1$, $F_{j_2}\to F_2$, $M_{F_{j_1}}\to
M_1$, $M_{F_{j_2}}\to M_2$, $F_{d_1}\to F_1^d$, $F_{d_2}\to
F_2^d$, $M_{F_{d_1}}\to M_1^d$, $M_{F_{d_2}}\to M_2^d$, while the
factors of the relaxation operator are expressed as follows:
 \be
    \label{Smain6F}
       \begin{array}{c}
   \dsp{ \Gamma_{j_1 j_2}(F_1 M_1,F_2 M_2)= \int \sum_{M_{d} \sigma \sigma^{'}}}
  \sum_{\lambda\pm 1}g_{j_1 d}^{{\bf k}\lambda}(F_1 M_1,F_d M_d) g_{d j_2}^{{\bf k}\lambda}(F_d M_d,F_2 M_2)\delta(\omega_{j_2 d}-\wk)
  \dsp{\frac{d {\bf k}}{(2\pi)^3}=} \\
   \dsp{=Q
   \sum_{F^d} (2F_d+1)(-1)^{2(I+1)-F_1-F_2-2F^d} W(J_{j_1} F_1 J_d F_d;I1)W(J_{j_2} F_2 J_d F_d;I1)
   \sum_{M_d \sigma,\sigma^{'}}C_{F_d M_d 1\sigma}^{F_1 M_1} C_{F_d M_d 1\sigma^{'}}^{F_2 M_2}}K^R(\sigma,\sigma^{'}),\\
    \dsp{\Gamma_{j_1j_2}(M_1 M_1^d ,M_2 M_2^d)= \int \sum_{\lambda\pm 1}g_{j_1 d}^{{\bf k}\lambda}(F_1 M_1,F_1^d M_1^d) g_{d j_2}^{{\bf k}\lambda}(F_2^d M_2^d,F_2 M_2)\delta(\omega_{j_2 d}-\wk)
  \frac{d {\bf k}}{(2\pi)^3}}=\\
  \dsp{=Q (2F_d+1)(-1)^{2(I+1)-F_1-F_2-F_1^d-F_2^d} W(J_{j_1} F_1 J_d F_1^d;I1)W(J_{j_2} F_2 J_d F_2^d;I1) C_{F_1^d M_1^d 1\sigma}^{F_1 M_1}
  C_{F_2^d M_2^d 1\sigma^{'}}^{F_2 M_2}K^R(\sigma,\sigma^{'}),} \\
   \end{array}
 \ee
 $$
  \dsp{ Q=\frac{2\parallel \mu_{j_1 d}\parallel\parallel \mu_{j_2 d}\parallel \omega_{j_2 d}^3}{\hbar
  c^3}}.
 $$
Let us now analyse the properties of factors $\Gamma_{j_1 j_2}(F_1
M_1,F_2  M_2)$.  Taking into account that Wigner functions are
orthogonal \rf{Ortog} and the property of the sums \rf{SummK} we
have:
 \be
  \label{Smain7F}
    \begin{array}{c}
  \dsp{\Gamma_{j_1 j_2}(F_1 M_1,F_2 M_2)= \frac{2Q}{3}\sum_{F^d} (2F_d+1)(-1)^{2(I+1)-F_1-F_2-2F_d} W(J_{j_1} F_1 J_d F_d;I1)W(J_{j_2} F_2 J_d
  F_d;I1)\delta_{F_1F_2}\delta_{M_1M_2}}.
  \end{array}
 \ee
  \end{widetext}
using the $6j$ representation for the Racah coefficients
\cite{Sobelman}:
 \be
   W(l_1 l_2 l_3 l_4;l_5 l_6)=(-1)^{-l_1-l_2-l_3-l_4}
   \left\{
   \begin{array}{ccc}
    l_1 & l_2 & l_5 \\
    l_4 & l_3 & l_6 \\
   \end{array} \right\},
 \ee
and their permutation symmetry as well as the property
 \cite{Sobelman}:
 \be
  \sum_l (2l+1)
  \left\{
   \begin{array}{ccc}
    l_1 & l_2 & l^{'} \\
    l_3 & l_4 & l \\
   \end{array} \right\}
   \left\{
   \begin{array}{ccc}
    l_3 & l_2 & l \\
    l_1 & l_4 & l^{''} \\
   \end{array} \right\}=\frac{\delta_{l^{'}l^{''}} }{2l^{''}+1},
 \ee
\\
we finally get:
 \be
  \label{Smain9F}
    \begin{array}{c}
  \dsp{\Gamma_{j_1 j_2}(F_1 M_1,F_2 M_2)= \frac{4\parallel \mu_{j_1 d}\parallel\parallel \mu_{j_2 d}\parallel
  \omega_{j_2 d}^3}{4 \hbar c^3(2J_{j_1}+1)} \times} \\
   \dsp{\times \delta_{J_{j_1} J_{j_2}}\delta_{F_1F_2}\delta_{M_1M_2}. }
  \end{array}
 \ee
Thus, the properties of the factors $\Gamma_{j_1 j_2}(F_1 M_1,F_2
 M_2)$ are absolutely similar to those discussed in Sec.
\ref{Spon}, which indicates that the operator describing
spontaneous transitions to the lower states is strictly diagonal.

\end{document}